\documentclass[aps,superscriptaddress, nofootinbib, showpacs]{revtex4}

\usepackage{devanagari}
\usepackage{dsfont}
\usepackage{amsmath,amssymb,amsfonts, bm, natbib}
\usepackage{dsfont}
\usepackage{epsfig}
\usepackage{graphicx}
\usepackage{slashed}
\usepackage{color}
\usepackage{tipa}

\usepackage{caption}
\usepackage{subcaption}
\usepackage{slashed}

\usepackage{commath}
\usepackage{calc}

\newcommand{\be}{\begin{equation}}
\newcommand{\ee}{\end{equation}}

\newcommand{\bea}{\begin{eqnarray}}
\newcommand{\eea}{\end{eqnarray}}

\newcommand{\bfig}{\begin{figure}}
\newcommand{\efig}{\end{figure}}

\usepackage{color,hyperref}
\definecolor{darkblue}{rgb}{0.0,0.0,0.3}
\hypersetup{colorlinks,breaklinks,
            linkcolor=darkblue,urlcolor=darkblue,
            anchorcolor=darkblue,citecolor=darkblue}

\providecommand*\url[1]{\href{#1}{#1}}
\renewcommand*\url[1]{\href{#1}{\texttt{#1}}}

\newcommand{\D}{{\cal D}}

\newcommand{\M}{{\cal M}}

\makeatletter
\newcommand*{\rom}[1]{\expandafter\@slowromancap\romannumeral #1@}
\makeatother

\begin{document}
\title{  	
Spontaneous CP breaking and number of fermionic families}
\author{Gauhar Abbas}
\email{gauhar.app@iitbhu.ac.in}
\affiliation{Department of Physics, Indian Institute of Technology (BHU), Varanasi 221005, India\footnote{Present address}, \\
Theoretical Physics
Division, Physical Research Laboratory, Navrangpura, Ahmedabad 380
009, India}
\begin{abstract}
We provide a theoretical justification for the existence of fourth family fermions of the Standard Model by showing that in a novel form of spontaneous $CP$ breaking, fourth family is inevitable.  This requires that fermions of third and fourth families  must be mirror matter conjectured long back by Lee and Yang. 
\end{abstract}


\maketitle
Mirror matter was first proposed by T.D. Lee and C. N. Yang  in thier seminal paper on parity violation\cite{Lee:1956qn}.  The idea was to restore parity which is a symmetry between left and right.    Since then various theoretical models based on mirror fermions are being proposed and investigated in literature \cite{Lee:1956qn,mirror1,mirror2,Lavoura:1997pq,Abbas:RRL,Abbas:LSLRRL,Abbas:LRSMM,Abbas:CP}.  The Large Hadron Collider (LHC) is actively looking for mirror matter\cite{Aad:2015tba}.

However, it is possible that nature is asymmetric with respect to parity, and still left-right symmetric when looked at through $CP$-mirror.  For instance, absence of right-handed particle can be compensated by right-handed anti-particle.   Having said that, we know that this is not even  respected by nature due to observed $CP$ violation.  However, amount of observed $CP$ very violation is small.  This gives us hope that  the observed small $CP$-violation in the SM could hint for a larger theory where  $CP$ is spontaneously broken.  Moreover, $CP$ violation  is one of the essential ingredient to explain matter-antimatter asymmetry of the Universe.  $CP$ violation is studied to a large degree in literature\cite{Branco:1999fs,Bigi:2000yz,Ivanov:2017dad}.

Apart from the so-called ``mirror matter", there is also possibility of an additional generation of ordinary quarks and leptons beyond the three generations of the fermions of the standard model(SM).  This framework is known as the SM4\cite{Frampton:1999xi,Holdom:2009rf,cline,BarShalom:2012ms}.  The SM4 is quite interesting from different theoretical perspectives.  For instance, it can address the hierarchy problem\cite{Bardeen:1989ds,Holdom:1986rn,King:1990he,Hill:1990ge,Hung:1996gj,Holdom:2006mr,Hung:2010xh}, and can explain origin of matter-antimatter asymmetry of the Universe\cite{Hou:2008ji,Ham:2004xh,Hou:2011df,Fok:2008yg}.  Furthermore, it can provide an explanation to flavour physics and CKM anomalies \cite{SAGMN08,SAGMN10,ajb10B,buras_charm,4Gflavor}.  It is quite interesting to note  that SM4 is still surviving in the scalar extensions of the SM\cite{Das:2017mnu}.

Fermionic mass hierarchy is another issue which is intriguing and fascinating.    This problem is so compelling that its solution is the only wish of Steven Weinberg to see in his life time\cite{CC:13oct17}.  However, this problem is very intricate in the sense that there are three class of hierarchies among quarks and leptons.   The first class of the mass hierarchy is among the fermionic families.  What we mean here is the mass hierarchy among fermionic families, ranging from the top quark of third family with a mass of order the electroweak scale to the electron mass of $0.511$ MeV.  The second class of mass hierarchy resides within the families, i.e., $m_d > m_u$,   $m_c > > m_s$,  $m_t >> m_b$. The third class of hierarchy resides in the quark-mixing angles.  Discovering a common simple and elegant explanation for all three hierarchies is one of the most challenging theoretical problems.  There have been several efforts in this direction\cite{tHooft:1971qjg,Weinberg:1971nd,Weinberg:1972ws,Georgi:1972hy,Mohapatra:1974wk,Barr:1976bk,Froggatt:1978nt,Cvetic:1985bg,
Balakrishna:1987qd,Davidson:1993xn,Haba:2002ek,Babu:1999me,Ross:2002fb,Giudice:2008uua,Bazzocchi:2008rz,King:2013hoa,
Ishimori:2014nxa,Hartmann:2014ppa,Diaz:2017eob}. A solution which can address the fermionic mass hierarchies among the three families and within the family along with quark-mixing is presented in Ref.~\cite{Abbas:2017vws}.

In this paper, we  propose a novel  form of spontaneous $CP$ breaking which predicts that minimum number of fermionic families to be four, and third and fourth families of the SM fermions must be mirror matter.   Furthermore, we obtain an explanation  for  mass hierarchy among the fermionic families.  This is achieved by introducing complex singlet scalar fields.

We begin with the theoretical possibility that any geometric symmetry of the Poincare group, for instance parity or charge conjugation, can be represented by a product  of an operator in external space and an operator in the internal space of symmetry group of the system \cite{Lee:1956qn,mirror1}.  This concept is extensively studied in literature\cite{mirror2,Lavoura:1997pq,Abbas:RRL,Abbas:LSLRRL,Abbas:LRSMM,Abbas:CP}.  

With this, we can define a novel and non-trivial $CP$ transformation for fermionic fields as adopted in Refs. \cite{Lavoura:1997pq,Abbas:CP}, and  given as,
\bea
\label{CP:trans1}
(CP) \psi_L  (CP)^\dagger   = \gamma^0 C \overline{\psi_L^\prime}^T,~~ (CP) \psi_R  (CP)^\dagger   = \gamma^0 C \overline{\psi_R^\prime}^T.
\eea
where $\psi_{L,R}^\prime$ are new fields after $CP$ transformation.  

Our main idea is to assume that fermionic families of the SM behave non-trivially under the tranformation defined in Eq.(\ref{CP:trans1}). For illustration, there are two possibilities,
\begin{enumerate}
\item
Third family is the $CP$ counter part of the first family, i.e. $(CP) \psi_L^1  (CP)^\dagger   = \gamma^0 C \overline{\psi_L^{3}}^T$ and $(CP) \psi_R^1  (CP)^\dagger   = \gamma^0 C \overline{\psi_R^{3}}^T$  where $\psi_{L,R}^1$ denotes left-handed doublet and right-handed singlet fermions of the first family, and similarly $\psi_{L,R}^3$ is the representative of the third family fermions.
\item
The other possibility is when second family is the  $CP$ counter part of the first family, i.e. $(CP) \psi_L^1  (CP)^\dagger   = \gamma^0 C \overline{\psi_L^{2}}^T$ and $(CP) \psi_R^1  (CP)^\dagger   = \gamma^0 C \overline{\psi_R^{2}}^T$ where $\psi_{L,R}^2$ denotes left-handed doublet and right-handed singlet fermions of the second family.
\end{enumerate}
In this work, we shall assume the first possibility which also explains, as we will discuss later,  mass hierarchy among fermionic families.  

With this assumption, we reach to an interesting conclusion that requirement of this novel spontaneous $CP$ violation  ensures that there must exist a fourth family which should be the $CP$ mirror-counter-part of the second family, i.e. we must have 
\be 
\label{CP:trans2}
(CP) \psi_L^2  (CP)^\dagger   = \gamma^0 C \overline{\psi_L^{4}}^T \
(CP) \psi_R^2  (CP)^\dagger   = \gamma^0 C \overline{\psi_R^{4}}^T,
\ee
where $\psi_{L,R}^4$ denotes the fourth family fermions.   

Thus, it is remarkable to note at this point that in this novel form of spontaneous $CP$ violation, fourth family of fermions should exist and must be mirror matter.  Moreover, third family of fermions are already discovered mirror fermions among them $\tau$ lepton was discovered in 1975, $b$ quark in 1977 and $t$ in 1995.  It is also concluded that if there exist more than four families of fermions in nature, every new family must accompany by its mirror counter-part.  This means number of families must be an even number in this new type of spontaneous $CP$ violation.

For elaboration of spontaneous $CP$ breaking in details, from here, we will assume the following  trivial $CP$ transformation:
\begin{eqnarray}
(CP)\mathcal{W^\mu} (CP)^\dagger &=& -  \mathcal{W}^\mu,
 (CP)\mathcal{B^\mu} (CP)^\dagger = -  \mathcal{B}^\mu, 
 (CP)\mathcal{G^\mu} (CP)^\dagger = -  \mathcal{G}^\mu
\end{eqnarray}
where $\mathcal{W}^\mu$ is the gauge field corresponding to the gauge group $SU(2)_L$, $ \mathcal{B}^\mu $ denotes the gauge field for symmetry $U(1)_Y$, and $ \mathcal{G}_{\mu}$ represents the gluon field.  Besides this, since it is confirmed now that neutrinos are massive particles, we extend the SM with three right-handed singlet neutrinos corresponding to four families.

Now we first delve into fermion masses and mass hierarchy among fermionic families. Once we assume $CP$ transformations in Eq.(\ref{CP:trans1}), the Yukawa operator cannot generate physical masses of fermions.  This is because that requirement of Eq.(\ref{CP:trans1}) forces the Yukawa couplings of first and third family to be  identical.  Hence, we assume that masses of fermions originate from dimension-5 operators.  For this purpose, we add  two complex singlet scalar fields, $ \text{ {\dn k}} _1$ and $\text{ {\dn k}} _2$, to the SM which transform under the SM symmetry $SU(3)_c \otimes SU(2)_L  \otimes U(1)_{Y} $ as\footnote{Consonant  letter ``\text{ {\dn k}}"(k\textschwa)  is taken from the  Devanagari script.  It is pronounced as ``Ka" in Kashmir\cite{Abbas:2017vws}.},
\begin{eqnarray}
 \text{ {\dn k}} _1:(1,1,0),~ \text{ {\dn k}} _2:(1,1,0).
 \end{eqnarray}

The  $CP$  transformation of doublet and singlets are given as,
\begin{eqnarray}
 (CP) \varphi (CP)^\dagger &=&  \varphi^{\dagger T},    (CP) \text{ {\dn k}} _1 (CP)^\dagger =   \text{ {\dn k}} _2^\dagger .
\end{eqnarray}

\begin{table}[h]
\begin{center}
\begin{tabular}{|c|c|c|}
  \hline
  Fields             &        $\mathcal{Z}_2$                    & $\mathcal{Z}_2^\prime$     \\
  \hline
  $\psi_R^1$                 &   +  &     -                                 \\
  $\psi_R^2$                 &   +  &     -                                   \\
  $\text{ {\dn k}} _1$                        & +  &      -                                                  \\
  $ \psi_R^3$     & -  &   +                                             \\
   $ \psi_R^4$     & -  &   +                                             \\
  $\text{ {\dn k}} _2$           & - &      +           \\
  \hline
     \end{tabular}
\end{center}
\caption{The transformations of right-handed fermions of different families and singlet scalar fields under symmetries  $\mathcal{Z}_2$,  and $\mathcal{Z}_2^\prime$ where  superscript is a family number.}
 \label{tab1}
\end{table} 
Moreover, the SM symmetry is further extended by imposing discrete symmetries  $\mathcal{Z}_2$ and $\mathcal{Z}_2^\prime$    on the right handed fermions of each family and scalar fields  $\text{ {\dn k}} _1$, and $\text{ {\dn k}} _2$ as shown in Table \ref{tab1}. Use of these discrete symmetries  was first discussed in Refs.\cite{Abbas:LSLRRL,Abbas:CP,Abbas:LRSMM}.

The Yukawa Lagrangian is completely forbidden by discrete symmetries  $\mathcal{Z}_2$ and $\mathcal{Z}_2^\prime$  now.   It is observed now that masses of  fermions of four families are given by dimension-5 operators through the following equation:
\bea
\label{mass1}
{\mathcal{L}}_{mass} &=& \dfrac{1}{\Lambda} \left[  \Gamma_1 \bar{\psi_L^1}  \varphi  \psi_R^1   \text{ {\dn k}} _1 +  \Gamma_1^{* } \bar{\psi^{3}_{L}}    \varphi  \psi_R^3  \text{ {\dn k}} _2 
+ \Gamma_2  \bar{\psi_L^1}  \tilde{\varphi} \psi_R^1   \text{ {\dn k}} _1  + \Gamma_2^{ *} \bar{\psi^{3}_{L}}  \tilde{\varphi}  \psi_R^3  \text{ {\dn k}} _2 \right]  \\ \nonumber
&+&  \dfrac{1}{\Lambda} \left[   \Gamma_3 \bar{\psi_L^2}  \varphi  \psi_R^2   \text{ {\dn k}} _1 +  \Gamma_3^{* } \bar{\psi^{4}_{L}}    \varphi \psi_R^4  \text{ {\dn k}} _2 
+  \Gamma_4  \bar{\psi_L^2}  \tilde{\varphi} \psi_R^2   \text{ {\dn k}} _1  + \Gamma_4^{ *} \bar{\psi^{4}_{L}}  \tilde{\varphi}  \psi_R^4  \text{ {\dn k}} _2 \right]  
+ \dfrac{c}{\Lambda} \bar{l_{L}^c}    \tilde{\varphi}^* \tilde{\varphi}^\dagger  l_L  
+  {\rm H.c.},
\eea
where superscripts shows the family number and $l_L$ denotes leptonic doublet of the SM.  

We note that mass hierarchy of fermionic families  is an outcome of the model discussed in this work.  For this purpose, we need to assume that vacuum expectation values (VEVs) of the complex singlet scalar fields are such that $ \langle \text{ {\dn k}} _2 \rangle >> \langle \text{ {\dn k}} _1 \rangle $, and couplings are such that $\Gamma_{3,4} >  \Gamma_{1,2} $.  This choice explains why second family fermions are heavier than those of the first family and, similarly  the reason that third family fermions are heavier than those of the second family.  This also establishes that fourth family must be heavier than the third family.  Masses of neutrinos are derived from the Weinberg operator.

For achieving an ultraviolet completion of this models given in table \ref{tab1}, we  introduce  atleast one vector-like isosinglet up quark, one vector-like isosinglet  down type quark, one isosinglet vector-like charged lepton, and one isosinglet vector-like neutrino.  Their transformation  under $SU(3)_c \otimes SU(2)_L   \otimes U(1)_{Y} $ is given by,
\begin{eqnarray}
Q &=& U_{L,R}:(3,1,\dfrac{4}{3});  D_{L,R}:(3,1,-\dfrac{2}{3}), \\ \nonumber
L &=&  E_{L,R}:(1,1,-2);  N_{L,R}:(1,1,0).
\end{eqnarray}

The mass Lagrangian for vector-like fermions is given by,
\begin{eqnarray}
\label{mass2}
\mathcal{L}_{V} &=& M_U \bar{U}_L U_R  + M_D \bar{D}_L D_R
+  M_E \bar{E}_L E_R  +  M_N \bar{N}_L N_R + {\rm H.c.}.
\end{eqnarray}
The interactions of vector-like fermions with the SM  fermions, for instance for quarks, are given by,
\begin{eqnarray}
\mathcal{L}_{Vff} &=& Y_1  \bar{q}_L^1 \varphi L_R  + Y_1^* \bar{q}_L^3 \varphi L_R +  Y_2  \bar{q}_L^1 \tilde{\varphi} L_R  + Y_2^* \bar{q}_L^3 \tilde{\varphi} L_R  
+ Y_3  \bar{q}_L^2 \varphi L_R  + Y_3^* \bar{q}_L^4 \varphi L_R  +  Y_4  \bar{q}_L^2 \tilde{\varphi} L_R  + Y_4^* \bar{q}_L^4 \tilde{\varphi} L_R 
 \\ \nonumber
&+&  \bar{Q}_L \Bigl( C_1  q_R^1  \text{ {\dn k}} _1 + C_1^* q_R^3  \text{ {\dn k}} _2   +  C_2  q_R^2  \text{ {\dn k}} _1 + C_2^* q_R^4  \text{ {\dn k}} _2 \Bigr)   +  {\rm H.c},
\end{eqnarray}
where $q_L$ is a quark doublet of the SM. 

The new physics which is entering in Eqs.(\ref{mass1})  and (\ref{mass2}) are vector-like fermions in our model.  These fermions are searched by the LHC, and the most recent searche excludes them approximately below 1 TeV\cite{Sirunyan:2018omb}.

The most general $CP$ invariant scalar potential of the model takes the following form:
\begin{eqnarray}
V &=& \mu  \varphi^\dagger \varphi    + \mu_1   \text{ {\dn k}} _1^{ \dagger}  \text{ {\dn k}} _1 +   \mu_2 \text{ {\dn k}} _2^{ \dagger}  \text{ {\dn k}} _2 
+  \lambda_1    (\varphi^\dagger \varphi)^2    
+ \lambda_2 \left[  ( \text{ {\dn k}} _1^{ \dagger}  \text{ {\dn k}} _1)^2  + \text{ {\dn k}} _2^{ \dagger}  \text{ {\dn k}} _2)^2 \right] \\ \nonumber
&+&  \Bigl[\mu_3 + \lambda_3  \varphi^\dagger \varphi 
+ \lambda_4 ( \text{ {\dn k}} _1^{ \dagger}  \text{ {\dn k}} _1 +   \text{ {\dn k}} _2^{ \dagger}  \text{ {\dn k}} _2)\Bigr] (\text{ {\dn k}} _1^2 + \text{ {\dn k}} _2^{\dagger 2})    
+ \lambda_{5} ( \text{ {\dn k}} _1^4 +  \text{ {\dn k}} _2^{\dagger 4})  + {\rm H.c.},
\end{eqnarray}
where we have introduced mass terms $\mu_1$ and $\mu_2$ which breaks symmetries $\mathcal{Z}_2$ and $\mathcal{Z}_2^\prime$ softly.

The vacuum expectation values(VEVs) after the spontaneous symmetry breaking(SSB) can be written as,
\bea
\langle \varphi \rangle &=& \dfrac{1}{\sqrt{2}}\begin{pmatrix} 0 \\ v \end{pmatrix}, \langle \text{ {\dn k}} _1 \rangle = \omega_1 e^{i \alpha_1}/\sqrt{2}, \langle \text{ {\dn k}} _2 \rangle = \omega_2 e^{i \alpha_2}/\sqrt{2}.
\eea 
For simplicity, we assume that parameters $\mu_3$, $\lambda_3$, $\lambda_4$, and $\lambda_5$ are real.  Furthermore, to show that spontaneous $CP$ breaking is  possible even when one of the VEVs of the singlet scalar fields is real, we assume that $\cos \alpha_2 =1$.

The scalar potential is minimized by solving  Eqs. $\dfrac{\partial V}{\partial v} =\dfrac{\partial V}{\partial \omega_1} = \dfrac{\partial V}{\partial \omega_2}  = \dfrac{\partial V}{\partial \cos \alpha_1}   = \dfrac{\partial V}{\partial \cos \alpha_2} =0$. For the case where $\cos \alpha_2 =1$, the minimum provides 
\be
\label{cos}
\cos  \alpha_1 = -\sqrt{\dfrac{\lambda_3 (8 \omega_1^2 - 9 \omega_2^2)-\sqrt{(\lambda_1 v^2 + 3 \lambda_3 \omega_2^2 +\mu)^2 - 8 \lambda_3^2 \omega_1^4}- 3 \lambda_3 (\lambda_1 v^2 + \mu)}{16 \omega_1^2 \lambda_3}}
\ee
The above equation in general breaks the $CP$ symmetry spontaneously.

Now we discuss the diagonalization fermionic mass matrices.  We can write the mass matrix for down type quarks approximately,
\begin{equation}
\label{mmd}
\begin{array}{ll|}
\M_\D = \left( \begin{array}{ccccc}
\dfrac{\Gamma_{11}^d v \omega_1 }{2 M}&   \dfrac{\Gamma_{12}^d v \omega_1 }{2 M} & \dfrac{\Gamma_{13}^d v \omega_2 }{2 M}&   \dfrac{\Gamma_{14}^d v  \omega_2 }{2 M}& \dfrac{1}{\sqrt{2}} v Y_1^d  \\
  \dfrac{\Gamma_{21}^d  v \omega_1 }{2 M}&   \dfrac{\Gamma_{22}^d v \omega_1 }{2 M} & \dfrac{\Gamma_{23}^d v v_6 }{2 M}& \dfrac{\Gamma_{24}^d v  \omega_2 }{2 M} & \dfrac{1}{\sqrt{2}} v Y_2^d\\
  \dfrac{\Gamma_{31}^d v \omega_1 }{2 M}&   \dfrac{\Gamma_{32}^d v \omega_1  }{2 M} & \dfrac{\Gamma_{33}^d v v_6 }{2 M} & \dfrac{\Gamma_{34}^d v  \omega_2 }{2 M} & \dfrac{1}{\sqrt{2}} v Y_3^d\\
   \dfrac{\Gamma_{41}^d v \omega_1 }{2 M}&   \dfrac{\Gamma_{42}^d v \omega_1  }{2 M} & \dfrac{\Gamma_{43}^d v v_6 }{2 M} & \dfrac{\Gamma_{44}^d v  \omega_2 }{2 M} & \dfrac{1}{\sqrt{2}} v Y_4^d\\
 \dfrac{1}{\sqrt{2}} \omega_1 C_1^d   &     \dfrac{1}{\sqrt{2}} \omega_1 C_2^d     &    \dfrac{1}{\sqrt{2}} \omega_2 C_3^d &   \dfrac{1}{\sqrt{2}} \omega_2 C_4^d     & M_{D}
\end{array} \right).
\end{array}
\end{equation}

  The mass matrix given in Eq.(\ref{mmd}) can be written into the following form:
\begin{equation}
\label{MD2}
\begin{array}{ll}
\M_\D = \left( \begin{array}{cc}
m_d&   p \\
  X  & M_{D}
\end{array} \right) \,, \qquad 
\end{array}
\end{equation}
where  the $3 \times 3$ block $m_d$ represents the  SM fermionic block and $M_{D}$ is  $2 \times 2$  block. 

The diagonalization of the mass matrix $\M_\D$ is done through the bi-unitary transformation,
\begin{equation}
\label{eqU}
\begin{array}{ll}
U^\dagger \M_\D V = \left( \begin{array}{cc}
\tilde{m}&   0 \\
  0 & \tilde{M}
\end{array} \right) \,, \qquad 
\end{array}
\end{equation}
where $\tilde{m} = \rm{diag} (m_d, m_s, m_b )$ and $\tilde{M} = \rm{diag} (m_d^\prime, M_D ) $.

We can diagonalize $ \M_{\D}^\dagger \M_\D$ through the matrix  $V$ which is given by,
\begin{equation}
\label{eqV}
\begin{array}{ll}
V  = \left( \begin{array}{cc}
K_d &   R \\
  S & T
\end{array} \right).
\end{array}
\end{equation}

The  following relations are obtained using Eqs.(\ref{eqU}) and (\ref{eqV}):
\begin{subequations}
\bea
\label{eqa1}
(m_d^\dagger m_d + X^\dagger X ) K_d + ( m_d^\dagger p + X^\dagger M_{D} ) S &=& K_d \tilde{m}^2, \qquad  \\ 
\label{eqb1}
(m_d^\dagger m_d + X^\dagger X ) R + ( m_d^\dagger p + X^\dagger M_{D} )T &=& R \tilde{M}^2, \\
\label{eqc1}
(p^\dagger m_d + M_D^\dagger X) K_d  + (p^\dagger p + M_{D^{1}}^\dagger M_{D}) S &=& S \tilde{m}^2, \\
\label{eqd1}
(p^\dagger m_d + M_D^\dagger X)  R + (p^\dagger p + M_{D^{1}}^\dagger M_{D}) T &=& T \tilde{M}^2.
\eea
\end{subequations}
In the limit $\tilde{M}^2 >> \tilde{m}^2$, we obtain from Eq.(\ref{eqc1}):
\be
S \simeq - (p^\dagger p + M_{D}^\dagger M_{D} )^{-1} (p^\dagger m_d + M_D^\dagger X) K_d.
\ee
Similarly we can obtain matrices $T$ and $R$.

We obtains from Eq.(\ref{eqa1})\cite{Botella:2016ibj},
\bea
K_d \mathcal{ H}_{eff} K_d^{-1}  = \tilde{m}^2,
\eea
where the squared matrix $\mathcal{ H}_{eff}$ is a Hermitian and given by,
\bea
\mathcal{ H}_{eff} &\simeq &  (m_d^\dagger m_d + X^\dagger X ) -  ( m_d^\dagger p + X^\dagger M_{D} ) 
  (p^\dagger p + M_{D}^\dagger M_{D})^{-1} (p^\dagger m_d + M_{D}^{\dagger} X).
\eea
Thus, using above matrix, we can derive matrix $K_d$. 

Vector-like quarks with charge assignments $Q= \dfrac{2}{3}$ or $Q = - \dfrac{1}{3}$ have been extensively studied in literature\cite{Branco:1986my,Lavoura:1992qd,Lavoura:1992np,Burgess:1993vc,delAguila:1998tp,AguilarSaavedra:2002kr,Cacciapaglia:2010vn,
Dawson:2012di,Aguilar-Saavedra:2013qpa,Ellis:2014dza,Angelescu:2015kga,Fajfer:2013wca,Alok:2015iha,Alok:2014yua,Chen:2017hak,Barducci:2017xtw,Barducci:2013zaa,Barducci:2014ila,Barducci:2014gna}.  
The Little Higgs models contain $Q= \dfrac{2}{3}$ isosinglet  charge assignment\cite{ArkaniHamed:2002qy,ArkaniHamed:2002qx}.  The isosinglet  charge assignment  $Q = - \dfrac{1}{3}$ appears in  $E_6$ GUTs\cite{Gursey:1975ki,Hewett:1988xc}.   However both charge assignments as well as vector-like leptons appear in the model discussed in this paper as well as in Refs.\cite{Abbas:LSLRRL,Abbas:CP,Abbas:LRSMM}.  Hence, a rich and novel phenomenology emerges out of these models.

Phenomenological data  will place bounds on the model discussed in this work.  For instance  the oblique parameters $S$, $T$ and $U$ may put stringent constraints\cite{Peskin:1990zt,Peskin:1991sw,Kennedy:1990ib,Kennedy:1991sn}.  The $S$, $T$ and $U$ parameters for an arbitrary number of families plus vector-like quarks can be found in Ref.\cite{Lavoura:1992np}.  For the model discussed in this paper, there will be additional CKM, FCNC  couplings appearing in the expressions of the $S$, $T$ and $U$ parameters which will make them relaxed with respect to electrweak precision data.  However, flavour physics data will provide more stringent constraints on this model. For instance, FCNC coupling   involving $s$ and $d$ quarks can be bounded by the process  $K^+ \rightarrow \pi^+ \nu \bar{\nu}$.

It will be interesting to comment briefly on  the so-called ``flavour anomalies" in quark-level $b \rightarrow s l \bar{l}$ transitions \cite{Aaij:2014ora,Aaij:2017vbb}.  In Ref.\cite{Botella:2017caf}, it is observed that  extending the SM by vector-like quarks and a heavy neutrino can provide an explanation for these anomalies.  Similar conclusion may also be obatained in the model discussed in this work.

In conclusion, we have  proposed a novel form of spontaneous $CP$ breaking which predicts the existence of the fourth family of the SM fermions.  Hence, this work presented in this paper,  for the first time,  provide a theoretical argument in the support of the existence of the fourth family of fermions of the SM.  For achieving this new form of the spontaneous $CP$ breaking, we have extended the SM by discrete symmetries $\mathcal{Z}_2$ and $\mathcal{Z}_2^\prime$.  Furthermore, in doing so, the masses of the SM fermions originate from  dimension-5 operators which are UV completed by vector-like fermions.  A detailed investigation of this model is a future goal.

\end{document}